\documentclass[twocolumn,aps,superscriptaddress,prl]{revtex4-2}
\usepackage{amsmath}
\usepackage{amsthm}
\usepackage{newtxmath}
\usepackage[utf8]{inputenc}
\usepackage{color}
\usepackage{float}
\usepackage{graphicx}
\usepackage{hyperref}
\usepackage{xcolor}
\hypersetup{
    colorlinks=true,
    linkcolor=blue,
    citecolor=blue,
    urlcolor=blue
}

\makeatletter

\usepackage{anyfontsize}

\makeatother

\begin{document}
\title{Disordered Parity Anomalous Semimetal}
\author{Shi-Hao Bi}
\address{Department of Physics, The University of Hong Kong, Pokfulam Road,
Hong Kong, China}
\author{Bo Fu}
\email{fubo@gbu.edu.cn}

\address{School of Sciences, Great Bay University, Dongguan 523000, China}
\author{Shun-Qing Shen}
\email{sshen@hku.hk}

\address{Department of Physics, The University of Hong Kong, Pokfulam Road,
Hong Kong, China}
\date{\today}
\begin{abstract}
The parity-anomalous semimetal (PAS) is a topological state of matter
exhibiting a semi-metallic nature and a half-quantized Hall conductance
of $e^{2}/h$ ($e$ is the elementary charge and $h$ is the Planck
constant). In this work, we investigate the disorder-driven topological
phase transition in a semi-magnetic narrow-gap band insulator thin
film. We demonstrate that strong disorder induces a transition from
the narrow-gap band insulator to a PAS phase, accompanied by the emergence
of a single gapless Dirac cone---the hallmark of the half-quantized
Hall effect. Calculations of the local density of states reveal the
spectral evolution underlying this transition, while finite-size scaling
of the real-space Hall conductivity confirms the robustness of the
half-quantized plateau over a finite range of disorder strengths.
Our findings establish disorder as a powerful tool for engineering
topological phases and provide new insights into the interplay between
topology and localization in quantum materials.
\end{abstract}
\maketitle
\textit{Introduction}---In ferromagnets, the anomalous Hall effect
and its quantized form are widely recognized, stemming from the combined
influence of spin-orbit coupling and ferromagnetism \citep{Nagaosa2010:RMP,Xiao2010:RPM,shen2012topological,Chang2023:RMP}.
While most previous research has centered on the quantum anomalous
Hall effect in gapped systems with integer topological invariants,
the systematic investigation of semi-metallic systems with half-integer
topological invariants has only been undertaken in recent years \citep{Fu2022:npjQM,Zou2022:PRB,Zou2023:PRB,Wang2024:PRB,Fu2024:NC,Fu2025:CommPhys}.
It was then reported that the measured Hall conductivity approaches
one-half in a semi-magnetic structure of Cr-doped topological insulator
$\mathrm{(Bi,Sb)_{2}\mathrm{Te}_{3}}$ \citep{Mogi2022:NatPhys}.
This experimental system constitutes a physical realization of PAS,
exhibiting a band structure with a single gapless surface Dirac cone
of electrons in the first Brillouin zone, thereby manifesting the
parity anomaly \citep{Qi2011:RMP,Redlich1984:PRD,Semenoff1984:PRL,Shen2024:coshare}
and leading to the half quantization of the Hall conductivity \citep{Fu2022:npjQM,Zou2022:PRB,Zou2023:PRB,Wang2024:PRB}.
Due to existence of a finite Fermi surface and nonzero longitudinal
conductivity, the PAS is apparently distinct from the quantum anomalous
Hall effect and fractional quantum anomalous Hall effect observed
in an insulating phase, which are characterized by the Chern numbers
and emergence of the chiral edge states \citep{Thouless1982:PRL,Haldane1988:PRL,Yu2010:Science,Chu2011:PRB,Chang2013:Science,Checkelsky2014:NatPhys,XXD2023:Nat1,XXD2023:Nat2,Lu2024:Nat}.
Consequently, there has been a significant research effort to comprehend
the origins of this effect, with numerous studies focusing on topics
such as the realization, robustness, and dissipative properties of
the half-quantized Hall effect \citep{Gong2023:NSR,Yang2023:CommPhys,Ning2023:PRB,Wang2024:PRB,Wan2024:PRB,Zhou2024:PRL}.

\begin{figure}[htbp]
\centering \includegraphics[width=8.5cm]{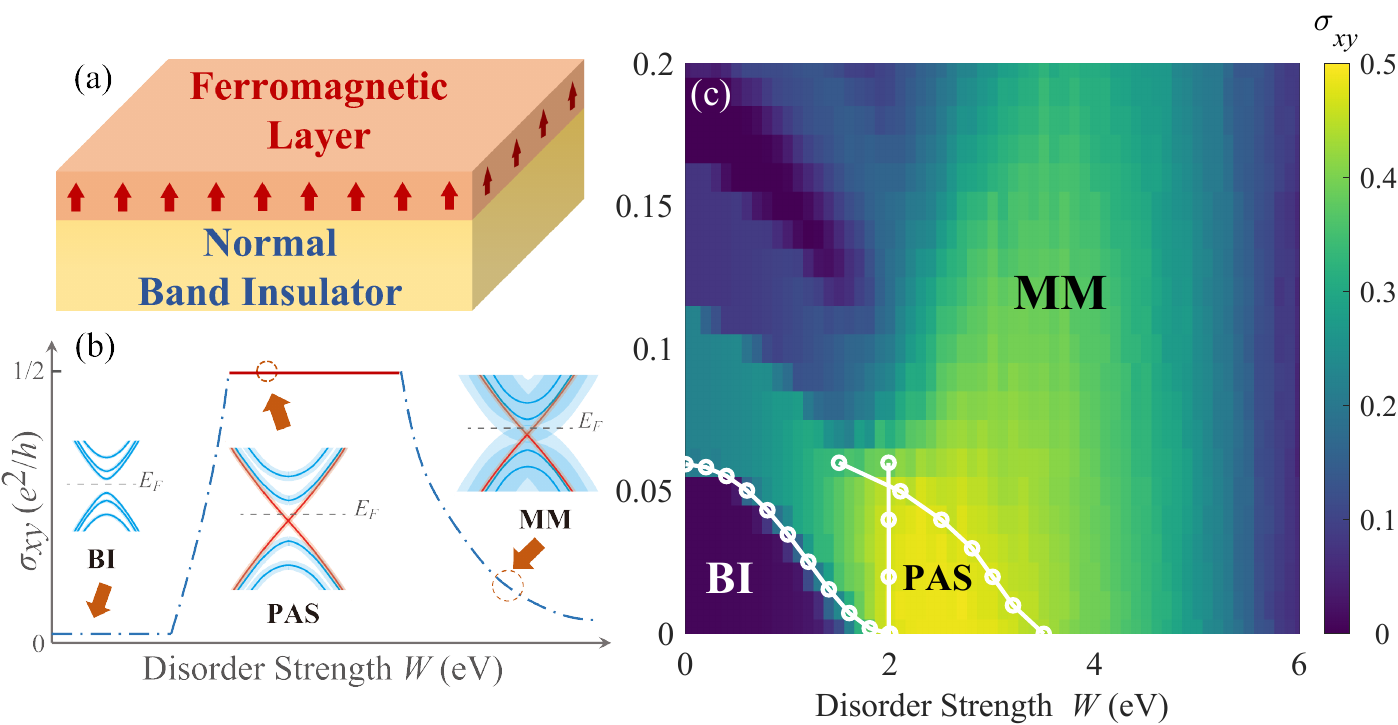} \caption{(a) Schematic diagram for a semi-magnetic structure of a narrow-gap
band insulator thin films. The red arrows indicate the alignment of
local magnetic moments. (b) Evolution of the Hall conductivity and
the quasi-particle spectrum via disorder. The color stripes represent
the energy broadening. (c) The phase diagram of the Hall conductivity
in the $W-E_{F}$ plane. The bright yellow areas highlight the PAS
phase, and the solid white circle lines indicate the phase boundaries
determined by means of the effective medium theory. Parameters used
are $L_{x}=L_{y}=20$, $L_{z}=10$, $L_{z}^{\mathrm{Mag}}=3$, and
lattice constants $a=b=1$ nm and $c=0.5$ nm. 50 random samples are
averaged for each point in the phase diagram. }
\label{fig:phase}
\end{figure}

Disorder, such as vacancies, defects, and impurities, is inevitable
in real materials and can induce remarkable phenomena in two dimensions
(2D) \citep{GoF1979:PRL,Mirlin2008:RMP,50YoAL}, including the metal-insulator
transition, quantum Hall effect \citep{Girvin2012:QHE}, and topological
Anderson insulator \citep{Li2009:PRL,Groth2009:PRL}. In addition
to driving the metal-insulator transition, disorder also plays a crucial
role in generating chiral edge states in topological phases \citep{Li2009:PRL,Groth2009:PRL,Stutzer2018:Nat,Meier2018:Science,Li2020:PRL,Liu2020:PRL,Zhang2021:PRL,Lin2022:NC,Dai2024:NatM,Ren2024:PRL}.
Therefore, grasping the influence of disorder on the stability and
emergence of PAS is paramount. In this work, we reveal that PAS is
not only stable against the disorder, but also can be precipitated
from a narrow-gap insulating phase through the very presence of disorder.
The phase diagram in Fig. \ref{fig:phase} is established by calculating
the Hall conductivity on a real space lattice numerically on a semi-magnetic
narrow-gap band insulator. The half quantized Hall conductance in
PAS is attributed to the emergence of a single gapless Dirac cone
induced by disorder, provided that the energy broadening does not
smear the gap between the gapless and massive Dirac cone. Furthermore,
an effective medium theory is developed to understand the formation
and breaking down of the PAS.

\textit{Model} \textit{and Hall conductivity}---Consider a semi-magnetic
structure of narrow-gap insulator film as shown in Fig. \ref{fig:phase}(a),
in which the magnetic ions are doped on the top layer of the film
to form a ferromagnetic layer.
The tight-binding model was introduced to describe the system \citep{shen2012topological},
\begin{equation}
H_{0}=\sum_{{\bf r}_{i}}\Psi_{{\bf r}_{i}}^{\dagger}M_{0}\Psi_{{\bf r}_{i}}+\sum_{{\bf r}_{i},\alpha=x,y,z}(\Psi_{{\bf r}_{i}}^{\dagger}\mathcal{T}_{\alpha}\Psi_{{\bf r}_{i}+{\bf e}_{\alpha}}+\mathrm{H.c.}),\label{eq:Ham}
\end{equation}
where $\mathcal{T_{\alpha}}=t_{\alpha}\tau_{z}\sigma_{0}-\frac{\mathrm{i}\lambda_{\alpha}}{2}\tau_{x}\sigma_{\alpha}$
and $M_{0}=\left(m_{0}-4t_{\parallel}-2t_{z}\right)\tau_{z}\sigma_{0}+V_{z}(i_{z})\tau_{0}\sigma_{z}$.
The default parameters are $\lambda_{x,y}=\lambda_{\parallel}=0.41$
eV, $\lambda_{z}=0.44$ eV, $t_{x,y}=t_{\parallel}=0.566$ eV, and
$t_{z}=0.40$ eV unless otherwise stated \citep{Zhang2009:NatPhys}. 
$\Psi_{{\bf r}_{i}}^{\dagger}$ and $\Psi_{{\bf r}_{i}}$ are four-component
creation and annihilation operators at site ${\bf r}_{i}$ encoding
both orbital (two-states) and spin degrees of freedom. $\tau_{\alpha}$
and $\sigma_{\alpha}$'s are the Pauli matrices acting on the orbital
and spin spaces, respectively. We propose the transition-metal pentatelluride
$\mathrm{ZrTe}_{5}$ as a candidate material, which is commonly regarded
as a weak topological insulator approaching the critical points for
a transition to a strong topological insulator \citep{Weng2014:PRX}.
Its low-energy states consist of four $p_{y}$ orbitals from two Te
atoms per unit cell, $|\mathrm{Te}_{1/2}p_{y}\uparrow\rangle$ and
$|\mathrm{Te}_{1/2}p_{y}\downarrow\rangle$ \citep{Chen2015:PRL}.
As the $\mathbb{Z}_{2}$ index is determined by $m_{0}$
and $t_{x,y,z}$ \citep{Kane2007:PRB,Volovik2010:JETP,Shen2011:SPIN,shen2012topological},
we model the narrow-gap trivial insulator by setting $m_{0}=-0.02$
eV, a value close to but on the opposite side of the transition relative
to $t_{z}$. Finally, the magnetic doping is modeled by introducing
a Zeeman potential $V_{z}(i_{z})$. $V_{z}(i_{z})=V_{0}=0.1$ eV for
the top layer $i_{z}\leqslant L_{z}^{\mathrm{Mag}}$, and 0 otherwise.
The case of a small but positive $m_{0}$ was used for a strong topological
insulator, which has been studied extensively \citep{shen2012topological}.

We then study the impact of disorder on the electrical Hall conductivity
of the semi-magnetic structure of narrow-gap band insulator thin film.
We introduce disorder through random on-site energies $u_{{\bf r}_{i}}$
which maintains the orbital-spin structure and are uniformly distributed
in $\left[-W/2,+W/2\right]$, leading to the impurity Hamiltonian
$H_{\mathrm{imp}}=\sum_{\mathbf{r}_{i}}\Psi_{{\bf r}_{i}}^{\dagger}u_{{\bf r}_{i}}\tau_{0}\sigma_{0}\Psi_{{\bf r}_{i}}$.
The Hall conductivity is computed numerically using the Prodan’s real-space 
non-commutative formula in Ref. \citep{Prodan2011:JPA}, which is usually used 
for calculating Chern number or quantized Hall conductivity in insulator \cite{Bellissard1994:JMP}:
\begin{equation}
\sigma_{xy}=\frac{e^{2}}{h}\left\langle 2\pi\mathrm{i}\mathrm{Tr}\left\{ P\left[-\mathrm{i}\left[x,P\right],-\mathrm{i}\left[y,P\right]\right]\right\} \right\rangle _{\mathrm{imp}},
\end{equation}
where $P$ denotes the projector onto the occupied states, and $x$
and $y$ are the coordinate operators. $\left\langle \cdots\right\rangle _{\mathrm{imp}}$
denotes the disorder-average. The applicability of the formula to the general case was discussed 
in Supplementary Material in Ref. \citep{Note-on-SI}. We take periodic boundary condition
in the $x$ and $y$ directions to eliminate the boundary effect,
and open boundary condition in the $z$ direction. The phase diagram
of the Hall conductivity, plotted as a function of disorder strength
$W$ and Fermi energy $E_{F}$, is depicted in Fig. \ref{fig:phase}(c).
Under weak disorder and at small Fermi energy (lower left of phase
diagram), the system remains in the BI phase with negligible Hall
conductivity, as indicated by the dark-purple color. When the Fermi
energy is elevated into regimes that populates electronic states near
the band edge of massive Dirac cones, non-vanishing Berry curvature
may arise. However, the resulting Hall conductance fails to achieve
quantization. On the contrary, with increasing disorder strength,
the PAS phase emerges strikingly, characterized by one half quantum
Hall conductivity $\frac{e^{2}}{2h}$. This phase persist over a finite
range of disorder, as indicated by the bright yellow region in the
phase diagram. When the disorder strength is even stronger, the Hall
conductivity deviates from the half quantized value, marking the emergence
of the marginal metal (MM) phase. Upon further increase in disorder,
$\sigma_{xy}$ eventually vanishes, and the system transitions into
an Anderson insulator (AI). In the MM phase, the Hall conductivity
continues to grow with sample size $L$ before saturating at a non-quantized
value. We identify this as a metallic phase because the Hall conductivity
must be quantized as an integer for an insulating state. This conclusion
is further supported by the calculations of the geometric mean of
density of states, as demonstrated in our subsequent analysis.

\begin{figure}[H]
\centering \includegraphics[width=8.5cm]{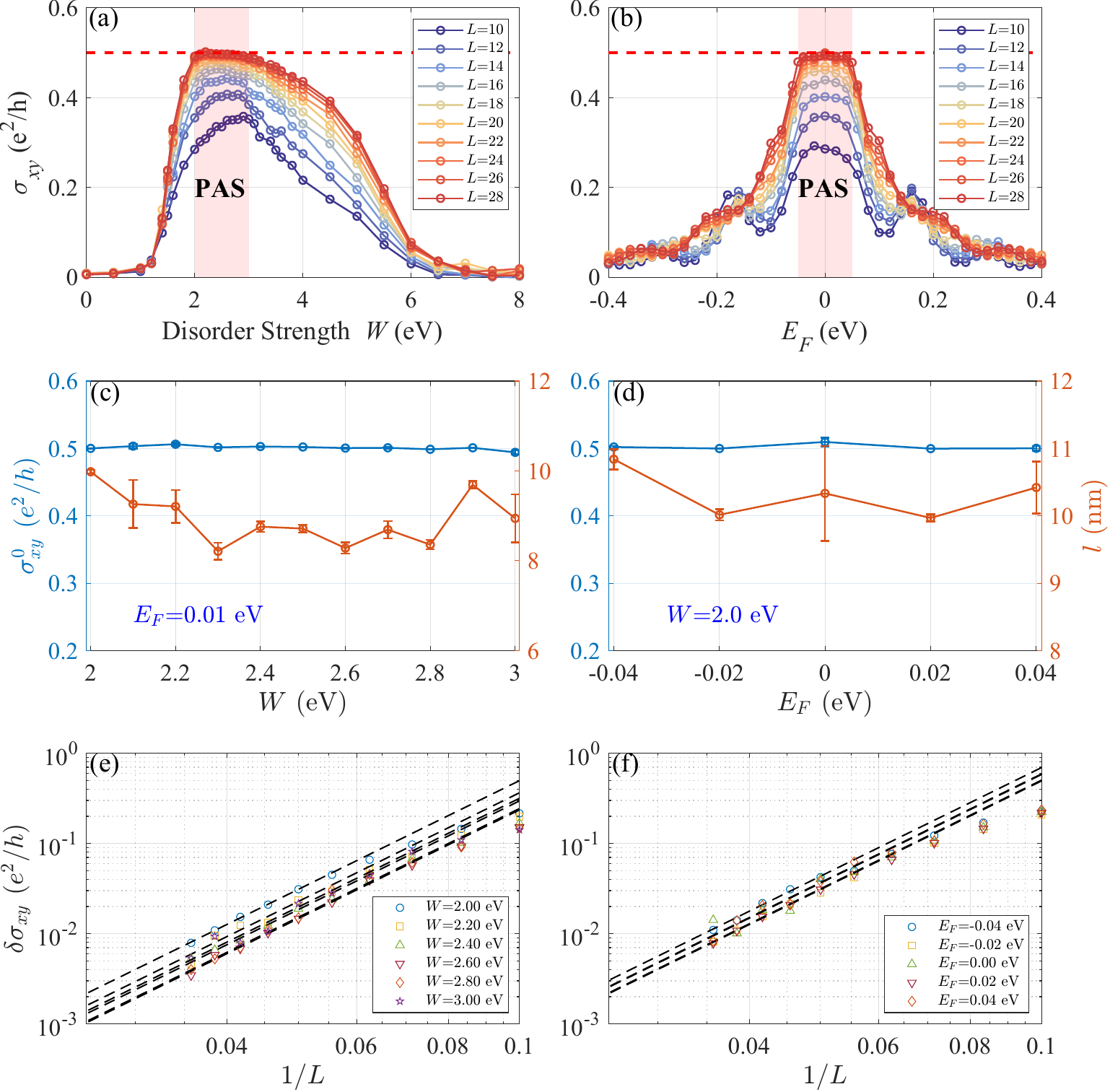} \caption{Real-space Hall conductivity as a function of (a) disorder strength
$W$ and (b) Fermi energy $E_{F}$ for varying system sizes. The finite-size
scaling analysis yields $\sigma_{xy}^{0}$ and $l$, which are exhibited
in (c) and (d), respectively. The corresponding quantization error
$\delta\sigma_{xy}$ is plotted in (e) and (f). The Hall conductivity
is averaged over 100 disordered samples for each point.}
\label{fig:fss}
\end{figure}

For a finite-sized system, the Hall conductance fails to attain precise
quantization within topologically nontrivial regimes. To elucidate
the disorder-driven topological phase transition, we engage in an
in-depth exploration of the finite-size effects of the real-space
Hall conductivity, opting to traverse the horizontal axis at $E_{F}=0.01$
eV and calibrating the vertical axis at $W=2.0$ eV, with the results
presented in Fig. \ref{fig:fss}. In Fig. \ref{fig:fss}(a), a pronounced
Hall conductance emerges within regions of moderate disorder, though
no quantized plateau is observed for smaller systems ($L=10$). As
the system size is enlarged, the Hall conductivity progressively converges
toward a half quantized plateau within the range $W\in\left[2.0,3.0\right]$
eV, as highlighted by the light red stripe. Analogously, Fig. \ref{fig:fss}(b)
illustrates that the Hall conductance, while exhibiting deviations
from quantization on smaller lattices, eventually manifests a half-quantized
Hall plateau for $\left|E_{F}\right|<0.05$ eV when increasing the
system size. 

We further analyze the finite-size scaling behavior of the PAS regime
to reveal the intrinsic topological characteristics of this novel
and exotic phase, and establish its robust quantization in the thermodynamic
limit. The Hall conductivity data exhibit a power-law scaling behavior,
which can be fitted with the following form:
\begin{equation}
\sigma_{xy}\left(L\right)=\sigma_{xy}^{0}\left[1-\left(l/L\right)^{4}\right],\label{eq:HallFSS}
\end{equation}
where $\sigma_{xy}^{0}$ represents the Hall conductivity in the thermodynamic
limit, and $l$ denotes a characteristic length scale beyond which
the Hall conductivity calculated on finite-size lattices asymptotically
approaches $\sigma_{xy}^{0}$. When the Fermi energy is fixed at $E_{F}=0.01$
eV (Fig. \ref{fig:fss}(c)), our finite-size scaling analysis yields
that $\sigma_{xy}^{0}$ fluctuates between $0.4942\pm2.96\times10^{-3}$
and $0.5063\pm2.29\times10^{-3}$, while $l$ is about $8.21\sim9.98$
nm. Meanwhile, at $W=2.0$ eV (Fig. \ref{fig:fss}(d)), the extrapolated
Hall conductivity converges to $0.4998\pm4.15\times10^{-4}\sim0.5096\pm6.33\times10^{-3}$,
with the characteristic length $l=9.97\sim10.84$ nm. Notably, the
largest system size in our numerical simulation is $28$ nm, which
outstrips the extracted $l$ by nearly threefold, thereby confirming
the reliability of our extrapolation and unambiguously verifying the
intrinsic topological nature of the PAS phase in the thermodynamic
limit. Finally, the quantization error $\delta\sigma_{xy}=\sigma_{xy}^{0}-\sigma_{xy}$
is presented as a log-log plot in Fig. \ref{fig:fss}(e,f), conclusively
demonstrating that the finite-size effects are significantly suppressed
as the system size increases, ultimately leading the Hall conductivities
to exhibit exact half-quantization as $L\to\infty$.

\textit{Local density of states}---The unexpected emergence of the
PAS phase from a topologically trivial narrow-gap band insulator prompts
an investigation into its intrinsic connection with gapless Dirac
physics. The quasi-particle picture is valid when the disorder strength
is far from reaching the Anderson transition point from MM to AI.
In this situation, the disorder only renormalizes the energy spectrum
of the quasi-particles and introduces a finite lifetime. Hence, by
examining the evolution of the spectral function with varying the
disorder strength, we can clearly observe the changes occurring during
the phase transitions. The spectral function $A(\epsilon,\mathbf{k})$
is defined by $A(\epsilon,{\bf k})=\sum_{n}\left\langle \psi_{n{\bf k}}\right|\delta\left(\epsilon-H\right)\left|\psi_{n{\bf k}}\right\rangle $,
where $\left|\psi_{n{\bf k}}\right\rangle =\frac{1}{\sqrt{S}}\left|u_{n{\bf k}}\right\rangle \mathrm{e}^{\mathrm{i}{\bf k}\cdot{\bf r}}$
are the Bloch states for the clean system with $H_{\mathrm{0}}\left|u_{n{\bf k}}\right\rangle =\epsilon_{n}\left|u_{n{\bf k}}\right\rangle $,
where $\epsilon_{n}$ is $n$-th energy eigenvalue. For a disordered
system, $A(\epsilon,\mathbf{k})$ can be numerically evaluated via
the Chebyshev polynomial expansion employing the standard kernel polynomial
method \citep{KPM2006:RMP,Fan2021:PhysRep}, 
\begin{equation}
A(\epsilon,{\bf k})=\frac{1}{\pi\epsilon_{\mathrm{max}}\sqrt{1-\widetilde{\epsilon}^{2}}}\sum_{n}\sum_{m=0}^{M-1}\alpha_{m}\left\langle \psi_{n{\bf k}}\right|T_{m}(\widetilde{H})\left|\psi_{n{\bf k}}\right\rangle ,\label{eq:SF}
\end{equation}
in which $\epsilon_{\mathrm{max}}$ is chosen as an adequately large
energy scale ensuring that $\widetilde{\epsilon}=\epsilon/\epsilon_{\mathrm{max}}$
and eigenvalues of $\widetilde{H}=H/\epsilon_{\mathrm{max}}$ lie
within the the domain of Chebyshev polynomials of the first kind $T_{m}(x)=\cos\left(m\arccos x\right)$.
$\alpha_{0}=g_{0}^{\mathrm{J}}$ and $\alpha_{m\geqslant1}=2g_{m}^{\mathrm{J}}T_{m}(\widetilde{\epsilon})$.
The Jackson damping kernel $g_{m}^{\mathrm{J}}$ serves to mitigate
the Gibbs oscillation arising from truncating the first $M$ terms
\citep{KPM2006:RMP,Fan2021:PhysRep}. 

\begin{figure*}[t]
\centering \includegraphics[width=18cm]{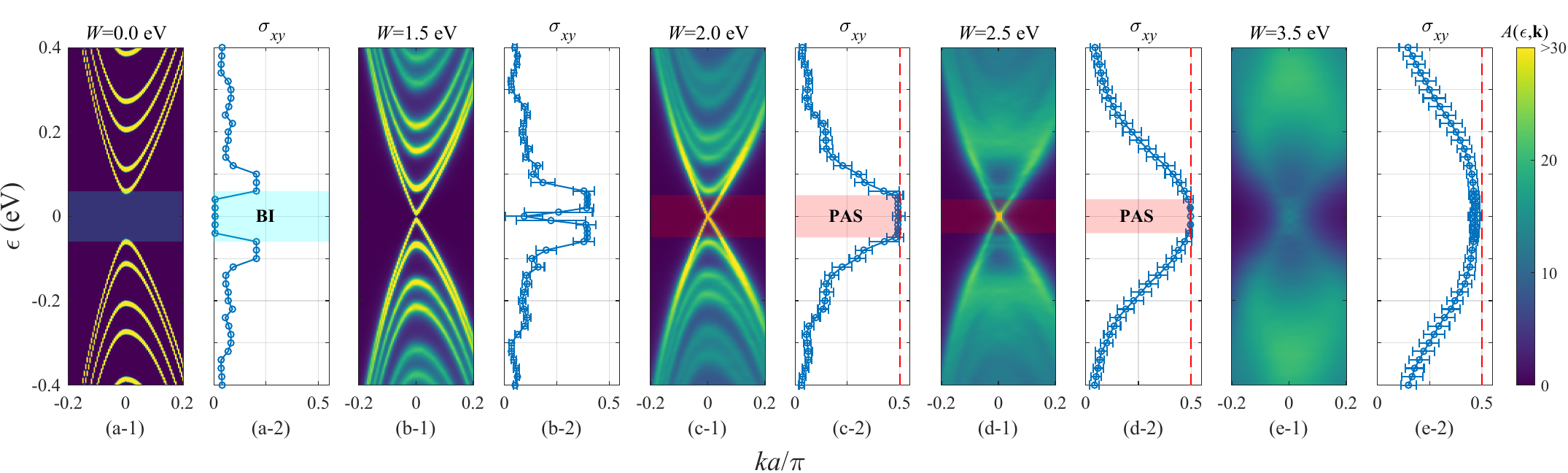} \caption{Evolution of the spectral function $A(\epsilon,\mathbf{k})$ with
the disorder strength $W$. The panels (a-e,1) are the spectral functions
for the disordered trivial band insulator, and panels (a-e,2) are
the corresponding real-space Hall conductivities calculated on a lattice
of size $L=28$. Each point of Hall conductance is averaged over 100
random samples. The cyan stripes indicate regions of trivial band
gaps wherein Hall conductivity vanishes, whereas the light red stripes
highlight renormalized Zeeman gap regions where it attains a PAS phase.
The lattice size used in the calculation of spectral functions: $L_{x}=L_{y}=400$,
$L_{z}=10$, and $L_{z}^{{\rm Mag}}=3$.We use $M=6000$ Chebyshev
moments to achieve high spectral resolution. The momentum path is
along the high-symmetry points $M-\Gamma-X$.}
\label{fig:LDOS}
\end{figure*}

The spectral functions for different disorder strengths are clearly
displaced in Fig. \ref{fig:LDOS}. Fig. \ref{fig:LDOS}(a-1) shows
that the trivial band structure has a finite gap at the $\Gamma$
point, and is a topologically trivial band insulator in the absence
of disorder. With increasing the disorder strength, the band gap progressively
shrinks (Fig. \ref{fig:LDOS}(b-1)) and ultimately closes near $W=2.0$
eV, giving rise to a single gapless Dirac cone as exhibited in Fig.
\ref{fig:phase}(c-1). Correspondingly, the Hall conductivity converges
to a half-quantized Hall plateau as demonstrated in Fig. \ref{fig:phase}(c-2).
The emergence of gapless Dirac cones constitutes a defining signature
of disorder-driven PAS phase, corroborated by precise calculations
of their Hall conductance. For even stronger disorder in Fig. \ref{fig:LDOS}(e-1)
, the gapless Dirac cone collapses as anticipated, and the Hall conductivity
deviates from half-quantization, giving rise to the MM. From a quasi-particle
perspective, our central result in this calculation reveals that disorder
can induce a single chiral gapless Dirac cone, and the PAS phase arises
when the chemical potential solely intersects the gapless Dirac cone
(despite broadening) within the broadened gapped bands.

\begin{figure}[htbp]
\centering \includegraphics[width=8cm]{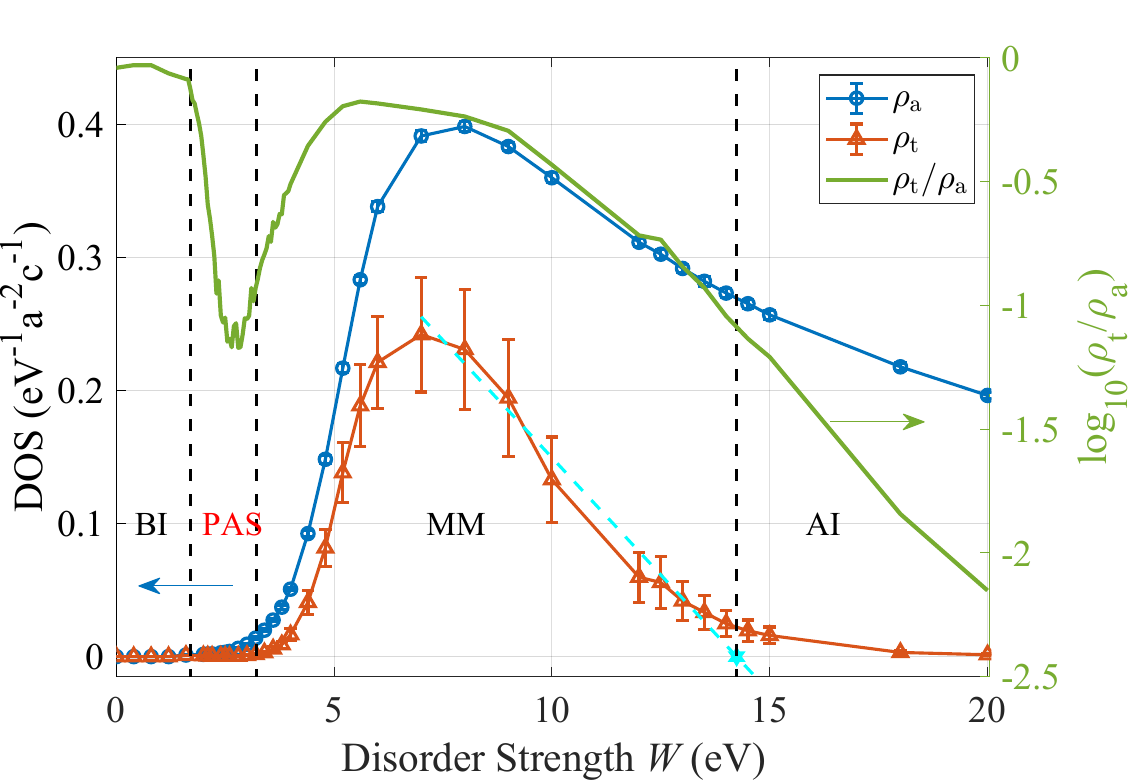} \caption{$\rho_{\mathrm{a}}(E)$ and $\rho_{\mathrm{t}}(E)$ versus disorder
strength at $E=0.01$ eV. The system size are the same as those in
Fig. \ref{fig:LDOS}. $M=6000$ Chebyshev moments and $N_{s}=30$
are used in the simulation. 20 samples are averaged for each point.
The black dashed lines at $W_{\mathrm{c},1}=1.99$ eV and $W_{\mathrm{c},2}=3.22$
eV indicate the transition points from BI to PAS and PAS to MM, as
determined by the effective medium theory. The cyan dashed line linearly
fitted to $\rho_{\mathrm{t}}$ intersects the horizontal axis at the
hexagonal marker ($W=14.24$ eV), thereby demarcating the Anderson
transition point.}
\label{fig:tDOS}
\end{figure}

These phase transitions can be characterized by analyzing the two
types of means of the local density of states (DOS): the arithmetic
mean $\rho_{\mathrm{a}}$ and the geometric mean $\rho_{\mathrm{t}}$.
The local DOS, $\rho_{{\bf r}}(\epsilon)=\sum_{\alpha}\left\langle {\bf r},\alpha\right|\delta\left(\epsilon-H\right)\left|{\bf r},\alpha\right\rangle $,
quantifies the amplitude of the wave function at site ${\bf r}$ for
a given energy $\epsilon$, where $\left|{\bf r},\alpha\right\rangle $
denotes an $\alpha$-orbital electron wave function at that site.
The spatial distribution of $\rho_{{\bf r}}(\epsilon)$ contains direct
information about the localization properties, which are closely intertwined
with the topology of the quantum system \citep{Pruisken,Tian2016:PRB,Zhang1994:PRL}.
The arithmetic and geometric mean DOS for a disordered system are
defined as $\rho_{\mathrm{a}}(\epsilon)=\left\langle \frac{1}{V}\sum_{i=1}^{V}\rho_{{\bf r}_{i}}(\epsilon)\right\rangle _{\mathrm{imp}}$
and $\rho_{\mathrm{t}}(\epsilon)=\exp\left[\frac{1}{N_{s}}\sum_{i=1}^{N_{s}}\left\langle \ln\rho_{{\bf r}_{i}}(\epsilon)\right\rangle _{\mathrm{imp}}\right]$,
respectively \citep{ZhangYY2012:PRB,ZhangYY2013:PRB}. Here we randomly
choose a finite number of lattice sites $N_{s}\ll V=L_{x}L_{y}L_{z}$
to improve the statistics of $\rho_{\mathrm{t}}$ \citep{Pixley2015:PRL}.
As illustrated in Fig. \ref{fig:tDOS}, by analyzing $\rho_{\mathrm{a}}$
and $\rho_{\mathrm{t}}$ at $E_{F}=0.01$ eV , we establish that the
narrow-gap band insulator film exhibits three quantum phase transitions:
BI$\to$PAS$\to$MM$\to$AI. Starting from the BI phase and considering
weak disorder, the presence of a finite band gap results in no states
within this gap, leading to $\rho_{{\rm t}}=\rho_{{\rm a}}=0$. Nevertheless,
the numerical calculation inevitably introduce a negligibly small
broadening parameter, which effectively introduces a uniformly distributed
nonzero local DOS in BI phase. This results in a ratio of $\rho_{{\rm t}}/\rho_{{\rm a}}\simeq$1
in this regime. As disorder increases, a phase transition from BI
to PAS occurs at $W_{\mathrm{c},1}=1.99$ eV, characterized by a strongly
suppressed $\rho_{{\rm t}}/\rho_{{\rm a}}$. In the PAS phase, the
emergence of gapless surface states---though metallic and extended
along the boundary---decays exponentially into the bulk with a characteristic
localization length $\xi$. Due to the localization of the surface
states, the ratio $\rho_{{\rm t}}/\rho_{{\rm a}}$ scales as ${\rm e}^{-L_{z}/\xi}$.
Hence, the decay of $\rho_{{\rm t}}/\rho_{{\rm a}}$ signals the emergence
of the surface states. This is consistent with our previous calculation
of Hall conductivity and spectral functions. Continuing to increase
the disorder strength leads to encountering two more quantum phase
transitions. First, the rise in the ratio $\rho_{{\rm t}}/\rho_{{\rm a}}$
indicates the formation of an extended metallic state in the bulk.
Upon further increasing disorder strength, the ratio's subsequent
collapse to zero provides a clear signature of the Anderson localization
transition, where all electronic states become exponentially localized.

\textit{Localization scenario of unitary class}---The
time-reversal symmetry breaking  induced by the magnetic layers assigns
the system to the unitary class, where all states are expected to
be localized except at critical points in 2D. Nonetheless, the time-reversal
symmetry breaking is inhomogeneous throughout the quasi-2D film and
confined to one surface, differing fundamentally from earlier study
with global symmetry breaking. The characteristic length scale for
time-reversal symmetry breaking across the entire film, $\ell_{\text{TRSB}}$,
grows exponentially with $L_{z}$ and exceeds any experimentally achievable
sample size. Based on these results, we characterize the PAS with 
half-integer quantum Hall effect as a stable phase rather than a critical point. 
We do not see any hint from the calculated data that the result is 
a finite size effect, which of course deserves further study.
A $\theta=\pi$ topological term in the unitary class was shown
to stabilize a metallic phase in single parameter scaling theory \citep{Ostrovsky2007:PRL},
offering independent support for our numerical findings.

\begin{figure}
\centering \includegraphics[width=8cm]{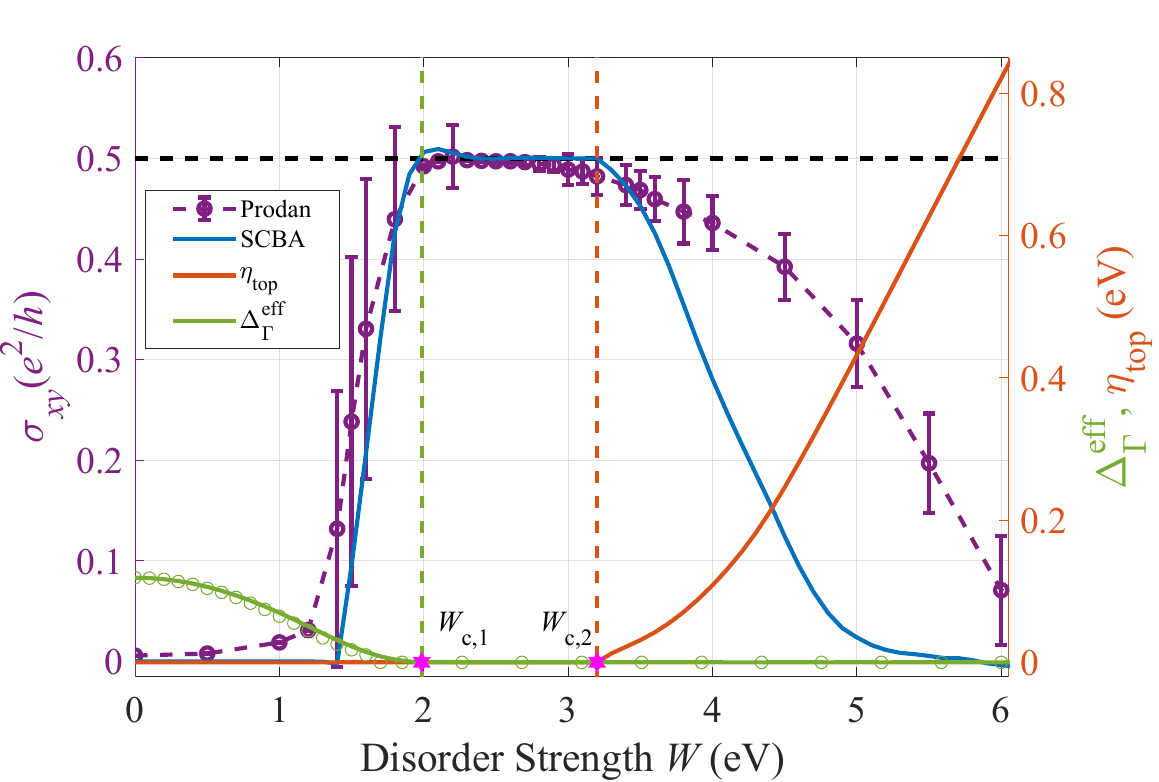} \caption{The Hall conductivity versus disorder strength $W$ for a narrow-gap
band insulator. Here we take $L_{x}=L_{y}=28$, and the data points
are averaged from 100 random samples. The blue solid line represents
the Hall conductivity, as calculated using the effective medium theory.
The green line indicates the effective band gap at $\Gamma$ point,
as evidenced by the green empty circles directly extracted from the
spectral function in Fig. \ref{fig:LDOS}. The orange line is the
energy broadening at the top layer ($i_{z}=1$). Two magenta pentagons
at $W_{\mathrm{c},1}$ and $W_{\mathrm{c},2}$ indicate the left and
right boundaries of the PAS phase.}
\label{fig:RSCN_SCBA_KB}
\end{figure}

\textit{Effective medium theory}---To further understand the origin
of the phase transitions from BI to PAS, we can employ the effective
medium theory in conjunction with the Kubo-Bastin formula for electrical
conductivity \citep{Note-on-SI,Bastin1971:JPCS,ChenYu2018:PRB,Bonbien2020:PRB}.
Within the framework of self-consistent Born approximation (SCBA),
the retarded self-energy is derived as \citep{Groth2009:PRL}
\begin{equation}
\Sigma^{R}\left({\bf k},E_{F}\right)=\sum_{{\bf k}'}\left\langle U_{{\bf k}{\bf k}'}\frac{1}{E_{F}-H_{0}({\bf k}')-\Sigma^{R}}U_{{\bf k}'{\bf k}}\right\rangle _{\mathrm{imp}}.
\end{equation}
Here, $U_{{\bf k}{\bf k}'}$ denotes the layer-resolved scattering
amplitude between planar momenta ${\bf k}$ and ${\bf k}'$ arising
from disorder, and $\Sigma^{R}$ is a $4L_{z}$-dimensional matrix
that incorporates layer, spin, and orbital indices. Numerical solution
shows the renormalized mass $\widetilde{m}_{0}\left(i_{z}\right)=m_{0}+\frac{1}{4}\mathrm{Tr}\left[\Sigma_{i_{z}i_{z}}^{R}(E_{F})\sigma_{0}\tau_{z}\right]$
exhibits insignificant layer dependence and is remarkably amplified
by disorder irrespective of its initial values. The evolution of the
quasi-particle band structure shown in Fig. \ref{fig:LDOS} is primarily
due to this effect. As shown in Fig. \ref{fig:RSCN_SCBA_KB}, the
green circles indicate the band gap extracted from the spectral function
in Fig. \ref{fig:LDOS}. The green solid line, representing results
from the effective medium theory, demonstrates good consistency with
these findings. When the gap closes at $W_{\mathrm{c},1}$, a half-quantized
Hall plateau emerges, as indicated by the vertical green dashed line.
Another key result is the pronounced layer dependence of the self-energy's
imaginary part \citep{Sarma2022:PRB}, $\eta(i_{z})=-\frac{1}{4}\mathrm{Im}\mathrm{Tr}\left[\Sigma_{i_{z}i_{z}}^{R}(E_{F})\right]$,
which governs energy level broadening. The band broadening $\eta_{\mathrm{bottom}}$
on the bottom surface is always present, regardless of disorder strength,
while the $\eta_{\mathrm{top}}$ on the top (the orange line in Fig.
\ref{fig:RSCN_SCBA_KB}) surges at a critical value $W_{\mathrm{c},2}\approx3.22$
eV of the disorder strength. The breakdown of quantized Hall conductivity
coincides with the emergence of $\eta_{\mathrm{top}}$ (indicated
by the vertical orange dashed line). Since $\eta_{\mathrm{top}}$
primarily affects the bulk-distributed higher-energy states of the
Dirac cone, it constitutes the key source of nonzero Hall conductivity
\citep{Zou2023:PRB}. The PAS-MM phase boundary (white dotted line
in Fig. \ref{fig:phase})) is defined by the emergence of $\eta_{\mathrm{top}}$,
which agrees well with the independent Hall conductivity simulations.
Therefore, nonzero $\eta_{\mathrm{top}}$ drives the PAS-to-MM transition.

\textit{Conclusion}---To conclude, the PAS is induced by disorder
in the semi-magnetic structure based on numerical calculation of the
Hall conductivity and local DOS, and can be understood very well in
the framework of the effective medium theory. We demonstrate that
strong disorder not only leads to a topological transition from the
BI to PAS phase, but also creates a single chiral gapless Dirac cone.
Our findings unequivocally illuminate the genesis of half-quantized
Hall phase in a topologically trivial narrow-gap band insulator film,
substantiates the paradigm of disorder-driven topological phase transition,
and provide a sturdy foundation for the exploration and development
of half-quantized Hall effects in quantum materials and devices.

\begin{acknowledgments}
We thank Dr. Huan-Wen Wang and Rui Chen for helpful discussions. This
work was supported by the Quantum Science Center of Guangdong-Hong
Kong-Macao Greater Bay Area (Grant No. GDZX2301005) and the Research
Grants Council, University Grants Committee, Hong Kong (Grants No.
C7012-21G and No. 17301823). B.F. is financially supported by Guangdong
Basic and Applied Basic Research Foundation No. 2024A1515010430 and
No. 2023A1515140008 and Guangdong Province Introduced Innovative R\&D
Team Program (Grant No. 2023QN10X136).
\end{acknowledgments}


%

\end{document}